# Using Harmonic Mean to Replace Tsallis' q-Average

Xiangjun Feng

*World Chinese Forum on Science of General Systems**

*Abstract*— In this paper, a unified mathematical expression for the constraints leading to the equilibrium distributions of both extensive and non-extensive systems is presented. Based on this expression, a recommendation is made to replace Tsallis' q-average without obvious physical meaning with the statistical harmonic mean for a generalized system.

## I. INTRODUCTION

There is no doubt that Tsallis entropy [1] is an important progress of modern statistical mechanics. However, in the final version, not the original one, of Tsallis' formalism, there is a weak point. The weak point is that the q-average's physical meaning is far from being obvious. Recently, H. Hasegawa [2] made a serious comment preferring Tsallis' original idea of the ordinary average. In previous work of the author [3], a unified mathematical expression about the constraints leading to the equilibrium distributions of non-extensive systems was given as the theorem 2. In this paper a generalized version of that expression is given. The generalized expression is with clear physical meaning. Based on the new expression, the standard Tsallis' power-law or so-called q-exponential has been found to be associated directly with a new constraint. The new constraint is related with the harmonic mean. Since the statistical harmonic mean is with a relatively clear physical meaning and can completely describe the new constraint, it is recommended to use it to replace Tsallis' q-average.

## II. THE UNIFIED MATHEMATICAL EXPRESSION FOR THE CONSTRAINTS LEADING TO EQUILIBRIUM DISTRIBUTIONS

In the previous work of the author [3], a unified mathematical expression for the constrains leading to the equilibrium distributions of non-extensive systems was given as the theorem 2. After a further and deep study, the author has generalized that unified expression. The generalized expression is described as follows.

*Theorem* If the dynamic probability distribution of $p = \{p_1, p_2, ...p_n\}$ of a generalized system is constrained by the following unified expression

$$\tfrac{1}{(q-1)}[p_1(f_1^{(q-1)} - 1) + p_2(f_2^{(q-1)} - 1) + ... + p_n(f_n^{(q-1)} - 1)]$$
$$= const,$$

(Eq.1)

it is possible for the probability distribution p to reach the equilibrium state of $f = \{f_1, f_2...f_n\}$. When $q \to 1$, the equation 1 becomes

$$p_1 \ln(f_1) + p_2 \ln(f_2) + ... + p_n \ln(f_n) = const,$$

where, the const is meant by a constant.

*Proof*

The Tsallis entropy can be expressed as follows

$$E = \tfrac{1}{(q-1)}(1 - p_1^q - p_2^q - ... - p_n^q),$$

and there is a natural constraint which can be express as

$$p_1 + p_2 + ... + p_n = 1.$$

Therefore, the corresponding Lagrangian without taking constants into consideration is

$$L = \frac{1}{(q-1)}(1 - p_1^q - p_2^q - ... - p_n^q)$$
$$+ \frac{m_1}{(q-1)}[p_1(f_1^{(q-1)} - 1) + p_2(f_2^{(q-1)} - 1) + ... + p_n(f_n^{(q-1)} - 1)]$$
$$+ m_2(p_1 + p_2 + ... + p_n),$$

where $m_1$ and $m_2$ are Lagrangian multipliers to be determined. In order to make L take the extreme, one has

$$\frac{\partial L}{\partial p_i} = 0$$

Since

$$\frac{\partial L}{\partial p_i} = \frac{-q}{(q-1)} p_i^{(q-1)} + \frac{m_1}{(q-1)}(f_i^{(q-1)} - 1) + m_2, i = 1,2,...n,$$

one can make $m_1 = q$ and $m_2 = \dfrac{q}{(q-1)}$, a zero $\dfrac{\partial L}{\partial p_i}$ will give

$$p_i = f_i, i = 1,2...,n.$$

The theorem has been proven.

## III. THE PHYSICAL MEANING OF THE UNIFIED MATHEMATICAL EXPRESSION OF THE CONSTRAINTS LEADING TO EQUILIBRIUM DISTRIBUTIONS

There are two types of laws about the stimulus-response mechanism of human being. One is Weber–Fechner law and the other is Stevens' power-law [4][5]. For the author, the equation 1 is related with both of them. Imagine each component of the distribution $f$ is a stimulus to a man or a woman. The response as a whole, per Stevens' power-law, will be something like

$$f^\alpha = \{f_1^\alpha, f_2^\alpha, ..., f_n^\alpha\}$$

The man or the woman will store this response $f^\alpha$ in his or her brain. Later on, when he or she "recalls" the distribution $f$, he or she will reconstruct the distribution $f$ under the constraint of that the statistical average of $f^\alpha$ will be a constant during the whole dynamic process of the reconstruction. Let $\alpha = q-1$ and take into consideration the natural constraint of

$$p_1 + p_2 + ... + p_n = 1,$$

one will understand that the constraint leading to the reconstructed equilibrium distribution $f$ is

$$p_1 f_1^{(q-1)} + p_2 f_2^{(q-1)} + ... + p_n f_n^{(q-1)} = const,$$

or,

$$\frac{1}{(q-1)}[p_1(f_1^{(q-1)} - 1) + p_2(f_2^{(q-1)} - 1) + ... + p_n(f_n^{(q-1)} - 1)] = const,$$

for a given $q$. That is exactly what being described in the above-mentioned theorem. It is amazing that nature follows exactly the same law as human being in the reconstruction of a distribution ! It should be mentioned that when $q \to 1$, the Stevens' power law can be replaced by Weber–Fechner law and the conclusion will remain unchanged.

## IV. THE NEW CONSTRAINT LEADING TO THE STANDARD FORM OF TSALLIS' POWER-LAW

The standard form of Tsallis' power law is [6]

$$f_k \propto [1 + \beta(q-1)E_k]^{\frac{-1}{(q-1)}}, k = 1, 2..., n,$$

where $\beta$ is a Lagrangian multiplier, $E_k$ is the energy of the $k$th component of the generalized system. For this particular distribution, per the equation 1, the corresponding constraint leading to this distribution is

$$p_1[1+\beta(q-1)E_1]^{-1} + p_2[1+\beta(q-1)E_2]^{-1} + ... + p_n[1+\beta(q-1)E_n]^{-1} = const$$

(Eq.2)

For a generalized system, let
$EF_i = 1 + \beta(q-1)E_i, i = 1, 2, ...n,$
one can define $EF_i$ as the effective energy of the $i$th component of the generalized system. From equation 2,

$$RH = p_1(\frac{1}{EF_1}) + p_2(\frac{1}{EF_2}) + ... + p_n(\frac{1}{EF_n}) = const$$

(Eq.3)

One has

$$H = \frac{1}{RH} = const \qquad (Eq.4)$$

It is obvious that $H$ is nothing else but the statistical harmonic mean of the effective energies of $EF_i, i = 1, 2, ...n$. The meaning of the equation 4 is that a constant statistical harmonic mean of the effective energies of a generalized system will lead to the standard form of Tsallis' power law. Compared with Tsallis'q-average, the statistical harmonic mean's physical meaning is much clearer. It is recommended that the Tsallis' q-average should be replaced by the statistical harmonic mean to make people feel much more comfortable when they use the Tsallis entropy for their own purposes.

## V. CONCLUSIONS AND DISCUSSION

A unified mathematical expression for the constraints leading to equilibrium distributions of both extensive and non-extensive systems is presented. The clear physical meaning of the expression is discussed. Based on the unified expression, It is recommended to use the statistical harmonic mean of effective energies to replace Tsallis' q-average.